\documentclass[sigconf]{acmart}

\usepackage{algorithm}
\usepackage{algpseudocode}

\usepackage{bm}
\usepackage{subcaption}
\usepackage{hyperref}

\AtBeginDocument{%
  }

\usepackage{multirow}  

\begin{document}

\title[SEGB]{SEGB: Self-Evolved Generative Bidding with Local Autoregressive Diffusion}

\author{Yulong Gao}
\email{gaoyulong8@jd.com}
\affiliation{%
  \institution{JD.com}
  \city{Beijing}
  \country{China}
}

\author{Wan Jiang}
\email{jiangwan1@jd.com}
\affiliation{%
  \institution{JD.com}
  \city{Beijing}
  \country{China}
}

\author{Mingzhe Cao}
\email{caomingzhe1@jd.com}
\affiliation{%
  \institution{JD.com}
  \city{Beijing}
  \country{China}
}

\author{Xuepu Wang}
\email{wangxuepu1@jd.com}
\affiliation{%
  \institution{JD.com}
  \city{Beijing}
  \country{China}
}

\author{Zeyu Pan}
\email{panzeyu3@jd.com}
\affiliation{%
  \institution{JD.com}
  \city{Beijing}
  \country{China}
}

\author{Haonan Yang}
\email{haonanyang@csu.edu.cn}
\affiliation{%
  \institution{JD.com}
  \city{Beijing}
  \country{China}
}
\author{Ye Liu}
\email{liuye162@jd.com}
\affiliation{%
  \institution{JD.com}
  \city{Beijing}
  \country{China}
}
\author{Xin Yang}
\email{yangxin81@jd.com}
\affiliation{%
  \institution{JD.com}
  \city{Beijing}
  \country{China}
}
\renewcommand{\shortauthors}{Gao et al.}

\begin{abstract}
In the realm of online advertising, automated bidding has become a pivotal tool, enabling advertisers to efficiently capture impression opportunities in real-time. Recently, generative auto-bidding has shown significant promise, offering innovative solutions for effective ad optimization. However, existing offline-trained generative policies lack the near-term foresight required for dynamic markets and usually depend on simulators or external experts for post-training improvement. To overcome these critical limitations, we propose Self-Evolved Generative Bidding (SEGB), a framework that plans proactively and refines itself entirely offline. SEGB first synthesizes plausible short-horizon future states to guide each bid, providing the agent with crucial, dynamic foresight. Crucially, it then performs value-guided policy refinement to iteratively discover superior strategies without any external intervention. This self-contained approach uniquely enables robust policy improvement from static data alone. Experiments on the AuctionNet benchmark and a large-scale A/B test validate our approach, demonstrating that SEGB significantly outperforms state-of-the-art baselines. In a large-scale online deployment, it delivered substantial business value, achieving a +10.19\% increase in target cost, proving the effectiveness of our advanced planning and evolution paradigm.
\end{abstract}

\ccsdesc[500]{Information systems~Computational advertising}

\keywords{Online Advertising, Auto-bidding, Autoregressive Diffusion Model, Offline Reinforcement Learning, Offline Policy Evolution}


\maketitle

\section{Introduction}

With the continuous advancement of digital commerce, online advertising platforms have grown significantly \cite{evans2009adsurvey, huh2024introduction}. Major platforms like Google \cite{google2021}, Alibaba \cite{alibaba2021}, and Facebook \cite{facebook2021} have developed automated bidding solutions. Advertisers specify marketing objectives and KPIs, while platforms leverage historical and real-time data to estimate CTR and CVR, automatically generating optimal bids. Given the fluid auction environments, auto-bidding is regarded as a long-horizon sequential decision-making process.

Reinforcement learning has emerged for bidding optimization \cite{cai2017rlad, he2021uscb, sutton2018reinforcement}. While traditional RL builds on Markov Decision Processes \cite{liu2023multi, ye2019deep, bellman1966dynamic}, recent research challenges this assumption \cite{aigb}, showing future states can depend on extended historical sequences. Various generative techniques have emerged in offline RL \cite{janner2021offline}: diffusion-based approaches \cite{ho2020ddpm, song2021scorebased} model long-range dependencies but can disrupt constraints when applied globally \cite{aigb}, while return-to-go methods like Decision Transformer \cite{chen2021dt} and variants \cite{janner2021offline, yamagata2023qgpo} lack mechanisms for planning with future context. Moreover, offline RL \cite{levine2020offline, prudencio2023survey} suffers from limited state-action coverage and restricted exploration beyond static datasets.

To address these challenges, we propose Self-Evolved Generative Bidding (SEGB), a synergistic offline framework. The "Self-Evolved" terminology emphasizes that the policy evolves entirely during offline training, reaching an improved state before deployment, rather than requiring online adaptation or continual learning. SEGB employs Local Autoregressive Diffusion (LAD) for high-fidelity state planning, providing future-aware context while respecting causal constraints. We integrate this foresight into a Decision Transformer, transforming it from a reactive imitator into a proactive planner. Finally, through Group Relative Policy Optimization (GRPO) \cite{shao2024deepseekmath}, the policy evolves entirely offline, discovering superior strategies beyond the dataset's limitations without requiring simulators or online interaction.

In summary, the main contributions of this framework can be summarized as follows:
\begin{itemize}
\item We introduce an end-to-end Self-Evolved Generative Bidding (SEGB) framework that synergistically combines Local Autoregressive Diffusion with future-state-aware reinforcement learning, uniquely enabling both high-fidelity causal planning and proactive decision-making for real-world dynamic budget and performance goals.
\item By incorporating a GRPO post-training fine-tuning strategy, the model evolves its policy entirely offline without simulators or online exploration, learning to discover superior strategies that transcend the limitations of the original dataset and mitigate distributional shift challenges.
\item After confirming its effectiveness in multiple offline experiments, we conducted online A/B tests on a real advertising platform. The results show that SEGB outperforms existing auto-bidding approaches while demonstrating strong practicality and scalability in large-scale deployment settings.
\end{itemize}
\vspace{-1mm}

\section{Preliminary}
\subsection{Problem Statement}

In online advertising, advertisers compete for impressions by submitting bids. During a period, $N$ opportunities arrive sequentially. The advertiser who submits the highest bid $b_i$ wins, obtaining value $v_i$ and incurring cost $c_i$. The objective is to maximize total value $\sum_i x_i v_i$ (where $x_i \in [0,1]$ indicates win probability) subject to budget constraint $\sum_i c_i x_i \leq B$ and KPI constraints $\frac{\sum_i c_{ij} x_i}{\sum_i p_{ij} x_i} \leq k_j, \forall j$. This yields the constrained optimization:

\begin{equation}
\begin{aligned}
& \text{maximize}   &\quad& \sum_{i} v_i x_i \\
& \text{s.t.} &    & \sum_{i} c_{i} x_i \leq B \\
&                   &    & \frac{\sum_i c_{ij} x_i}{\sum_i p_{ij} x_i} \leq k_j \quad ,\forall j  \\
&                   &    & 0 \leq x_i \leq 1 \quad\quad,\forall i \\
\end{aligned}
\end{equation}

The optimal bid is \cite{he2021uscb}: $bid_i^* = \lambda_0 v_i + k_i\sum_{j=1}^{J} \lambda_j p_{ij}$, where $\lambda_j$ are Lagrange multipliers for constraints.

\subsection{Auto-Bidding as Sequential Decision-Making}
Due to dynamic auction environments, bidding parameters must be adjusted in real time, formulating auto-bidding as a sequential decision problem. At each time $t$, an agent observes state $s_t$ (budget, delivery time, impressions, costs, CPC/CPA), selects action $a_t = (a_t^{\lambda_0}, \ldots, a_t^{\lambda_J})$ via policy $\pi$, and receives reward $r_t$. A trajectory $\tau$ is a sequence of states, actions, and rewards over the campaign duration.

\section{THE SEGB PARADIGM FOR AUTO-BIDDING}
\label{sec:paradigm}
To bridge the critical offline-to-online gap in auto-bidding, we conceptualize our solution, SEGB, not as a monolithic model, but as a synergistic, multi-stage paradigm. This paradigm systematically addresses the core challenges of planning, decision-making, and exploration within a fully offline setting. Each stage builds upon the last, forming a logical progression from understanding the future to acting intelligently and finally, to evolving beyond the initial data.

\paragraph{High-Fidelity Trajectory Planning}
The process begins with tackling the planning deficit. We employ a Local Autoregressive Diffusion (LAD) model for state trajectory generation. Unlike standard diffusion models that can violate causal constraints, LAD autoregressively predicts each future state based on its historical context. This ensures the generation of high-fidelity, causally consistent trajectories that serve as a realistic "sandbox" for downstream decision-making.

\paragraph{Foresight-driven Action Generation}
With a reliable plan of the future, we then focus on generating proactive actions. We evolve the standard Decision Transformer (DT) by integrating the predicted next-state information from our LAD planner. By conditioning on both the long-term goal (return-to-go) and a concrete, immediate future state, our Next-State-Aware DT transitions from a reactive imitator to a forward-looking agent, producing more accurate and strategically sound actions.

\paragraph{Offline Policy Evolution}
The final and most crucial stage addresses the exploration dilemma, enabling the policy to self-evolve entirely from the static offline dataset. To break free from the limitations of the data, we fine-tune the policy using Group Relative Policy Optimization (GRPO). Guided by a robustly trained IQL critic, this offline evolution step allows the policy to safely explore and adopt superior strategies that may not have been explicitly demonstrated in the original trajectories. This elevates the agent from merely imitating good behavior to actively discovering better bidding policies.

Together, these three interconnected stages form the SEGB paradigm: a synergistic framework that combines high-fidelity planning, foresight-driven decisions, and offline policy evolution to deliver superior and robust bidding performance.
\section{The SEGB Methodology}

In this section, we present the detailed technical implementation of the Self-Evolved Generative Bidding (SEGB) framework. As outlined in our paradigm (Section~\ref{sec:paradigm}), the methodology is structured around three synergistic stages: High-Fidelity State Planning, Foresight-driven Action Generation, and Offline Policy Evolution. We will now elaborate on the models and algorithms employed in each of these stages.

\begin{figure*}[t]
    \centering
    \includegraphics[width=0.95\textwidth]{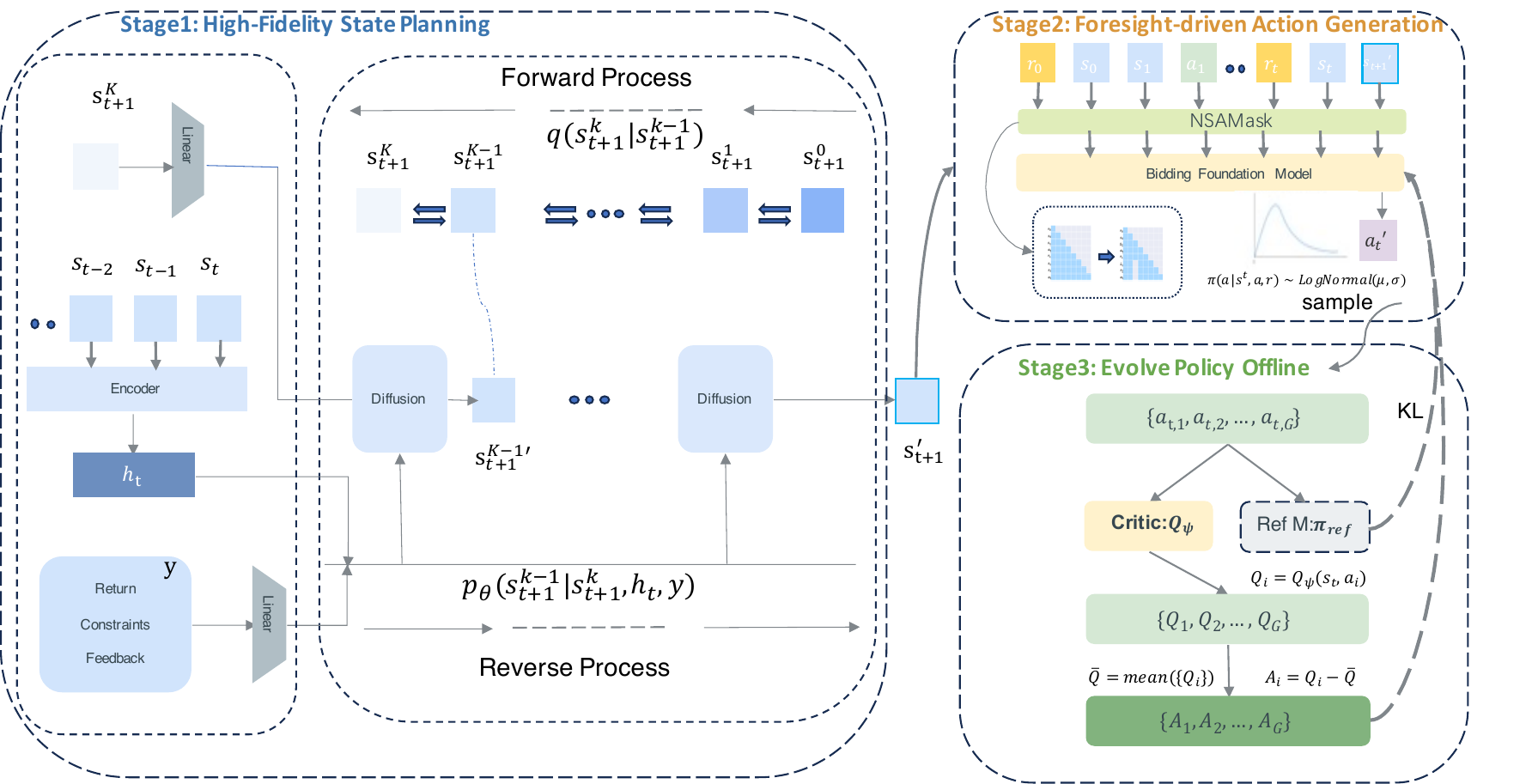}
    \caption{\small Figure 1: Overview of the SEGB Framework. SEGB consists of three stages. (1) Planning: A LAD model generates a high-fidelity future state prediction $(s'_{t+1})$. (2) Action Generation: A Next-State-Aware DT conditions on this prediction to generate an action $a'_t$. (3) Offline Evolution: The policy is then evolved via GRPO, guided by a frozen Critic and Reference Model to update the DT. Note that Stage 3 is only performed during offline training; online inference relies solely on the efficient Stage 1 and Stage 2 pipeline.}
    \Description{The SEGB framework.}
    \label{fig:segb_framework}
\end{figure*}
\vspace{-3mm}

\subsection{High-Fidelity State Planning via Local Autoregressive Diffusion}

In practical auto-bidding scenarios, the state sequence (e.g., remaining budget, accumulated conversions, elapsed time) exhibits strong temporal correlations and must respect specific real-world constraints, such as a monotonically decreasing budget. Traditional diffusion models (e.g., DDPM \cite{ho2020ddpm}) typically operate on entire data samples in a "global" fashion, which can overlook or violate time-dependent constraints. Moreover, global generation makes it difficult to enforce causal dependencies—the future should depend on the past, not vice versa. To address this challenge, we introduce \textbf{Local Autoregressive Diffusion (LAD)}, which generates each future state locally and autoregressively, conditioned on historical context. This design ensures high-fidelity trajectory synthesis that preserves both temporal causality and domain-specific constraints.


\subsubsection{LAD: Model Formulation and Framework}

The core idea behind LAD is to model the state generation process autoregressively, thereby preserving temporal dependencies. Formally, we aim to maximize the likelihood of observing the state trajectories in our dataset $D$. For a given trajectory $\tau = (s_1, s_2, \dots, s_T)$, this is expressed as:

\begin{equation}
\label{eq:mle}
\max_\theta \mathbb{E}_{\tau \sim D}\left[\prod_{t=1}^T p_{\theta}(s_t | s_{<t}, y(\tau))\right]
\end{equation}
where $y(\tau)$ represents campaign-level conditional attributes. This formulation naturally captures the requirement that each state $s_t$ depends on its history $s_{<t}$.

To implement this, LAD applies the diffusion and denoising processes locally to each state $s_t$ while conditioning on the historical context. The reverse denoising process, which generates the state, is conditioned on an embedding of the history, $z_t = f(s_1, \dots, s_{t-1})$:
\begin{equation}
p_{\theta}(s_t^{k-1}|s_1,\dots,s_{t-1},s_t^k,y(\tau)) = p_{\theta}(s_t^{k-1}|z_t, s^{k}_t, y(\tau))
\end{equation}
This ensures that the generation of each state explicitly accounts for the preceding states, enforcing causal adherence throughout the trajectory.

\subsubsection{Forward and Reverse Diffusion Processes}
\paragraph{Forward Process.}
The forward process for each local state $s^0_t$ follows a standard Markovian diffusion, gradually adding Gaussian noise over $K$ steps according to a predefined variance schedule $\{\beta_k\}_{k=1}^K$. This process is detailed in foundational works such as DDPM \cite{ho2020ddpm}. The distribution of a noisy state $s_t^k$ given the original state $s^{k-1}_t$ is:
\begin{equation}
q(s^k_t \mid s^{k-1}_t) = \mathcal{N}(s^k_t; \sqrt{1 - \beta_k} s^{k-1}_t, \beta_k I)
\label{eq:add_noise}
\end{equation}
Using the reparameterization trick, we can directly sample a noisy state $s^k_t$ at any step $k$ from the original state $s^0_t$:
\begin{equation}
s^k_t = \sqrt{\bar{\alpha}_k} s^0_t + \sqrt{1 - \bar{\alpha}_k}\epsilon, \text{where } \epsilon \sim \mathcal{N}(0, I), \alpha_k = 1 - \beta_k, \text{ and } \bar{\alpha}_k = \prod_{i=1}^{k}\alpha_i.
\end{equation}

\paragraph{Reverse Process.}
The reverse process learns to denoise $s^k_t$ back to $s^{k-1}_t$ in an autoregressive manner. It estimates the noise $\epsilon$ using a neural network $\epsilon_\theta$ that is conditioned on the noise level $k$, the historical context embedding $z_t$, and the campaign attributes $y(\tau)$. It is trained by minimizing the simplified objective from DDPM \cite{ho2020ddpm}, which is the mean squared error between the true and predicted noise:
\begin{equation}
\mathcal{L}_{\text{LAD}} = \mathbb{E}_{(\tau, s_t, \epsilon, k)} \left[ \left| \epsilon - \epsilon_\theta (s^k_t, k, z_t, y(\tau)) \right|^2 \right]
\label{eq:ladloss}
\end{equation}

Once trained, this network is used to iteratively generate a clean state. At each reverse step, we first compute the noise prediction $\hat{\epsilon}_k$ using classifier-free guidance for improved conditional generation:

\begin{equation}
\hat{\epsilon}_k = \epsilon_\theta(s^k_t, z_t, k) + \omega \left(\epsilon_\theta(s^k_t, z_t, y(\tau), k) - \epsilon_\theta(s^k_t, z_t, k)\right)
\end{equation}
where $\omega$ is the guidance strength.  This prediction is then used to compute the mean $\mu_{\theta}$ of the posterior distribution $p_{\theta}(s^{k-1}_t|z_t, s^k_t, y(\tau))$ from which the denoised state $s^{k-1}_t$ is sampled:
\begin{equation}
\mu_{\theta}(s^k_t, z_t, y(\tau), k) = \frac{1}{\sqrt{\alpha_k}}\left(s^k_t - \frac{\beta_k}{\sqrt{1 - \bar{\alpha}_k}}\hat{\epsilon}_k\right)
\end{equation}
During inference, we start with pure noise $s^K_t \sim N(0, I)$ and iteratively apply this reverse step K times to generate the clean state prediction $s_0^t$.

The output of this planning component is a high-fidelity prediction of the next state, $\hat{s}_{t+1}$. This crucial piece of forward-looking information serves as a key input to our decision-making agent, as detailed in the next section.


\subsection{Foresight-driven Action Generation with a Next-State-Aware DT}

With a high-fidelity state planner in place, the next step in our framework is to develop a decision-making agent that can effectively leverage this foresight. Standard sequential models like the Decision Transformer (DT) \cite{chen2021dt} are inherently reactive; they learn to act based on past events and a long-term, often sparse, `Return-to-Go` (RTG) signal. This purely backward-looking approach is a critical limitation in auto-bidding. An agent that only knows its past and its ultimate goal (the final RTG) is effectively driving in the dark, unable to make tactical adjustments based on predictable short-term outcomes, such as impending budget exhaustion.

To overcome this fundamental limitation, our second innovation is to evolve the DT into a \textbf{Next-State-Aware} agent. We explicitly incorporate the predicted next state $\hat{s}_{t+1}$—provided by our LAD component—into the DT's decision-making context. This represents a paradigm shift from reactive imitation to proactive, foresight-driven planning. Unlike RTG, which provides only an abstract long-term goal, the explicit next state offers a concrete, immediate target, enabling the agent to make tactically sound decisions with a clear short-term objective in mind.

\subsubsection{Model Formulation and Dual-Signal Guidance}
The standard DT \cite{chen2021dt} conditions its action $a_t$ on a history of states, actions, and returns-to-go: $\tau = (R_1, s_1, a_1, \dots, R_t, s_t)$, where $R_t = \sum_{i=t}^{T} \gamma^{i-t} r_i$. The policy is thus modeled as $\pi_{\text{DT}}(s_{\leq t}, a_{<t}, R_{\leq t})$.

In our approach, we adopt an inverse dynamics-inspired modeling strategy by explicitly incorporating the predicted next state $\hat{s}_{t+1}$ as a forward-looking signal. Specifically, we transform the sequential decision-making process into a predictive modeling framework where the policy is conditioned on partial future state information:
\begin{equation}
\label{eq:ns_dt}
a_t \sim \pi_{\theta}(a | s_{\leq t}, a_{<t}, R_{\leq t}, \hat{s}_{t+1})
\end{equation}

The necessity and innovativeness of this formulation lie in its creation of a dual-signal guided learning strategy.** By conditioning on both a long-term, strategic objective (the return-to-go, $R_t$) and a short-term, tactical objective (the immediate next state, $\hat{s}_{t+1}$), we provide a much richer and more stable learning signal.

This is particularly effective in auto-bidding for two reasons. \textbf{First}, bidding rewards (conversions) are extremely sparse, making the long-term $R_t$ signal weak and often delayed. The predicted next state, $\hat{s}_{t+1}$ (e.g., predicted remaining budget, predicted number of clicks), provides a dense, immediate, and concrete target for the agent to aim for at every step, drastically improving learning stability. \textbf{Second}, it allows the agent to reason about crucial constraints. For example, by seeing that the predicted next state $\hat{s}_{t+1}$ is approaching a budget limit, the agent can learn to proactively reduce its bid, a sophisticated behavior that is difficult to learn from a sparse final reward signal alone.

This forward-looking design provides the model with an explicit mechanism to anticipate how the environment will evolve in response to its action. Consequently, the agent is better equipped to generate stable and purposeful action predictions, bridging the gap between backward-looking inputs and a proactive, future-conditioned policy.

\subsubsection{Architecture and Supervised Pre-training}
The backbone of our agent is a GPT-like Transformer architecture that employs causal self-attention. States, actions, returns, and predicted states are first projected into a shared embedding space. These token embeddings, along with positional encodings, are then fed into the Transformer. The model outputs a distribution over the action space from which we can sample $a_t$.

The initial training of this Next-State-Aware DT is conducted in a supervised manner on the offline dataset. We aim to learn a high-quality initial policy $\pi_\theta$ by minimizing the behavioral cloning loss, which for continuous actions is typically the Mean Squared Error (MSE) between the predicted action and the ground-truth action from the dataset:
\begin{equation}
\mathcal{L}_{\text{DT}}(\theta) = \mathbb{E}_{(\tau, \hat{s}) \sim D} \left[ \left( a_t - \pi_{\theta}(s_{\leq t}, a_{<t}, R_{\leq t}, \hat{s}_{t+1}) \right)^2 \right]
\label{eq:dtloss}
\end{equation}
This pre-training phase yields a strong, foresight-driven policy. However, this policy is still constrained by the quality of the offline data. It serves as a high-quality starting point for the final offline policy evolution stage, which aims to improve upon these learned behaviors.

\subsection{Offline Policy Evolution}
\label{sec:evolution}

A policy pre-trained via supervised learning, even one enhanced with foresight, is fundamentally constrained by the quality and coverage of the offline dataset. It can replicate successful past behaviors, but it cannot discover novel, potentially superior strategies. To create a truly intelligent bidding agent, we must enable the policy to evolve beyond the training data. This is the final and most critical component of the SEGB framework, designed to solve the \textbf{exploration dilemma} and elevate the policy beyond the confines of its training data using only offline methods.

This offline policy evolution is achieved through a two-step offline optimization process: first, training a reliable critic to guide the evolution, and second, using this critic to fine-tune the policy itself.

\subsubsection{IQL-based Critic Training}
To guide the policy optimization, a reliable Q-function (critic) is required to estimate the value of state-action pairs. We train this critic, using Implicit Q-Learning (IQL) \cite{Kostrikov2021iql}. IQL is particularly well-suited for offline settings as it avoids the explicit evaluation of out-of-distribution (OOD) actions by learning the Q-function via expectile regression. To capture policy-specific context, our critic is a Transformer that conditions its value estimates on the historical trajectory:
\begin{equation}
Q_{\phi}^{\pi}(s_{t}, a_{t}) = \mathrm{QT}{\phi}(s_{t}, a_{t}; s_{<t}, a_{<t})
\end{equation}
The IQL training objective involves minimizing the expectile loss between the Q-values and target values derived from a separate state-value network $V_\psi(s)$:
\begin{equation}
\mathcal{L}_{\text{V}}(\psi) = \mathbb{E}_{(s,a) \sim D} \left[ L_2^\tau(Q_\phi(s,a) - V_\psi(s))\right]
\end{equation}
\begin{equation}
\mathcal{L}_{\text{Q}}(\phi) = \mathbb{E}_{(s,a) \sim D} \left[(Q_\phi(s,a) - (r(s,a) + \gamma V_\psi(s')))^2 \right]
\end{equation}
\begin{equation}
    \label{eq:iql_loss}
\mathcal{L}_{\text{IQL}}(\phi, \psi) = \mathcal{L}_Q(\phi) + \mathcal{L}_V(\psi)
\end{equation}
where $L_2^\tau$ is the expectile loss function. This process yields a robust critic, , ready to guide the evolution of our bidding policy.

\subsubsection{Offline Policy Evolution with GRPO}
With the IQL-trained critic $Q_{\phi}$ in place, the pivotal final step is to refine the bidding policy $\pi_{\theta}$ to surpass the quality of the offline dataset. This requires an algorithm capable of safe and stable policy improvement using only static data.

\paragraph{Motivation: Why a Hybrid IQL-GRPO Approach?}
Before describing the GRPO fine-tuning process, we first clarify our design rationale, as this hybrid approach differs from standard applications of either method alone.

Our approach is not standard GRPO, but a purpose-designed hybrid for offline policy learning in complex POMDPs. The synergy addresses critical limitations of either method in isolation:

\begin{itemize}
    \item \textbf{IQL Alone is Insufficient}: While IQL learns robust Q-functions via expectile regression~\cite{Kostrikov2021iql}, direct policy extraction remains challenging. Standard extraction methods—such as greedy argmax or Advantage-Weighted Regression (AWR)—are unstable in continuous, high-dimensional action spaces. Empirically, as shown in Table~\ref{tab:overall}, IQL alone achieves only 325.89 on AuctionNet (100\% budget), significantly underperforming even the standard DT baseline (335.34). This confirms that value learning alone is insufficient without a stable policy optimization mechanism.
    
    \item \textbf{GRPO Alone is Infeasible}: GRPO~\cite{grpo} typically requires reliable value signals, often obtained from rule-based reward models or accurate environment simulators. However, our bidding environment—characterized by black-box auction dynamics and highly volatile traffic patterns—precludes accurate simulation. Any simulator would suffer from compounding errors over the long (48-step) bidding horizon, making pure GRPO optimization unreliable.
    
    \item \textbf{Our Synergy}: By combining IQL and GRPO, we leverage the strengths of both. IQL provides stable, offline value estimation without any environment interaction or simulator; GRPO then performs guided, stable policy optimization using these value estimates. Critically, unlike online methods like PPO~\cite{ppo} which require on-policy rollouts from a live environment, we use IQL's offline-trained Q-function as a fixed value oracle. This eliminates the need for online data collection while maintaining stable policy gradients through GRPO's clipping and KL-penalty mechanisms.
\end{itemize}

Empirically, Table~\ref{tab:ablation} validates this design. The ``w/o GRPO'' variant (LAD + NSA-DT only) achieves 346.4, while the full model reaches 355.99, demonstrating that GRPO contributes +9.59 (+2.77\%) beyond a strong, foresight-aware baseline. Compared to the IQL baseline (325.89), our complete pipeline achieves +30.10 (+9.2\%), confirming that this hybrid approach is essential for effective offline policy evolution in our setting.

\paragraph{Why GRPO? The Rationale for Offline Policy Gradient.}
After the initial supervised pre-training, we fine-tune the policy to elevate it beyond simple imitation. Our choice of Group Relative Policy Optimization (GRPO) \cite{grpo} is motivated by comparing it with alternative fine-tuning paradigms:
\begin{itemize}
    \item \textbf{Supervised Fine-tuning (on expert data)}: Continuing to train with a Mean Squared Error loss, even on the best trajectories, only refines imitation. It cannot synthesize novel, superior actions because it lacks an understanding of the action-value landscape.
    \item \textbf{Online Policy Gradient} (e.g., PPO \cite{ppo}): These methods require a live, interactive environment to collect on-policy data for stable gradient updates, rendering them unsuitable for our offline setting.
    \item \textbf{Preference-based Optimization} (e.g., DPO \cite{rafailov2023dpo}): Emerging from the field of Reinforcement Learning from Human Feedback (RLHF) \cite{rlhf_christiano, rlhf_instructgpt}, methods like DPO are state-of-the-art for model alignment. However, they rely on explicit preference data (e.g., "action A is better than B"), which is unavailable in standard offline RL logs, and creating it from trajectories can be complex and biased.
\end{itemize}
GRPO provides a more direct and suitable solution. It uses our IQL-trained critic as a proxy for preference, directly optimizing the policy for higher estimated returns from standard trajectory data. This approach effectively balances policy improvement with the stability required in an offline context, making it the ideal choice for evolving our bidding strategy beyond the dataset's limitations.

\paragraph{GRPO Post-training.}
With the IQL critic $Q_\phi$ providing reliable value estimates, the final stage is to evolve the policy $\pi_\theta$ beyond the dataset's limitations. We employ Group Relative Policy Optimization (GRPO), using the pre-trained policy $\pi_{\theta_{\text{old}}}$ as the initial reference. GRPO maximizes an objective guided by the advantage estimates $\hat{A}_i$, which are derived from our critic $Q_{\phi}$.

The objective is constructed modularly. First, we define the importance sampling ratio $r_i(\theta)$:
\begin{equation}
    r_i(\theta) = \frac{\pi_{\theta}(o_i \mid q)}{\pi_{\theta_{\text{old}}}(o_i \mid q)}
\end{equation}
This ratio is used in the PPO-style clipped surrogate objective, $\mathcal{L}^{\text{CLIP}}$, which ensures stable policy updates:
\begin{equation}
    \label{eq:grpo_clip}
    \mathcal{L}^{\text{CLIP}}_i(\theta) = \min \Big( r_i(\theta) \hat{A}_i, \, \text{clip}\big(r_i(\theta), 1-\varepsilon, 1+\varepsilon\big) \hat{A}_i \Big)
\end{equation}
Finally, the full per-sample objective, $\mathcal{L}^{\text{GRPO}}$, incorporates a KL-divergence penalty to regularize the policy and prevent it from deviating too far from a trusted reference policy $\pi_{\text{ref}}$:
\begin{equation}
    \label{eq:grpo_loss}
    \mathcal{L}^{\text{GRPO}}(\dots) = \frac{1}{G} \sum_{i=1}^{G} \left( \mathcal{L}^{\text{CLIP}}_i(\theta) - \beta \cdot \mathbb{D}_{KL} \big[ \pi_{\theta}(\cdot \mid q) \parallel \pi_{\text{ref}}(\cdot \mid q) \big] \right)
\end{equation}

\paragraph{Implementation Note.} In practice, optimizing this joint loss is often best achieved via a structured two-step process for stability: we first train the planner to convergence by minimizing $\mathcal{L}_{\text{LAD}}$, and then, with the planner frozen, train the policy by minimizing $\mathcal{L}_{\text{DT}}$. This practical approach ensures that the policy is trained with a high-quality, stable foresight signal.

\subsection{SEGB Training}
The training of SEGB follows a powerful two-stage paradigm: supervised pre-training to build a strong policy foundation, followed by offline reinforcement learning for policy evolution.
In the first stage, we focus entirely on supervised learning to create a foresight-aware initial policy. This is guided by a unified objective, $\mathcal{L}_{\text{supervised}}$, which combines the diffusion loss for the planner and the behavioral cloning loss for the policy:
\begin{equation}
\label{eq:supervised_loss}
\mathcal{L}_{\text{supervised}} = \mathcal{L}_{\text{LAD}} + \mathcal{L}_{\text{DT}}
\end{equation}
Optimizing this objective endows the agent with a strong ability to imitate expert behavior based on future plans, all without any value estimation.
The second stage enables the agent to evolve beyond simple imitation. This reinforcement learning phase begins by preparing a reliable value guide: we first train a robust Q-function critic by minimizing the IQL expectile loss, $\mathcal{L}_{\text{IQL}}$. Then, with this pre-trained critic frozen to provide stable value estimates, the policy from Stage 1 is fine-tuned by maximizing the GRPO policy improvement objective, $J_{\text{GRPO-bid}}$. This two-step "prepare guide, then evolve" process allows SEGB to safely discover superior strategies in a fully offline manner. The entire structured procedure is summarized in Algorithm \ref{algo:segb_training}.

\begin{algorithm}[H]
\small
\caption{Training of SEGB}
\label{algo:segb_training}
\begin{algorithmic}[1]
    \Require Randomly initialized planner $\theta_{\text{LAD}}$, policy $\theta_{\pi}$, critic $\phi, \psi$; bidding trajectory dataset $\mathcal{D}$.
    \Ensure Optimized policy $\theta_{\pi}^{\text{final}}$.
    
    \While{not converged on supervised objectives}
        \State Sample a batch of trajectories $\mathcal{B}$ from $\mathcal{D}$;
        \For{\textbf{all} $\tau \in \mathcal{B}$}
        \State Sample $k\sim\text{Uniform}(1,K)$, $\epsilon\sim\mathcal{N}(0,I)$;
        \State Compute ${x}_k(\tau)$ via $q({x}_k(\tau)|{x}_0(\tau))$ in Eq~(\ref{eq:add_noise});
        \State Compute $\mathcal{L}(\theta_{LAD},\theta_{\pi})$ by Eq~(\ref{eq:supervised_loss});
        \State Perform gradient descent to optimize $\theta_{LAD}$ and $\theta_{\pi}$;
        \EndFor
    \EndWhile
    \State Let $\theta_{\pi}^{\text{pre-trained}} \gets \theta_{\pi}$;
    
    \Statex
    \While{not converged on IQL objective}
        \State Sample a batch of trajectories $\mathcal{B}$ from $\mathcal{D}$;
        \For{\textbf{all} $\tau \in \mathcal{B}$}
        \State Compute $\mathcal{L}(\theta_{LAD},\theta_{\pi})$ by Eq~(\ref{eq:iql_loss});
        \State Perform gradient descent to optimize $\phi$ and $\psi$;
        \EndFor
    \EndWhile
    \State Freeze critic $\phi, \psi$;
    
    \While{not converged on GRPO objective}
        \State Sample a batch of trajectories $\mathcal{B}$ from $\mathcal{D}$;
        \For{\textbf{all} $\tau \in \mathcal{B}$}
        \State Compute policy evolution objective $J_{\text{GRPO}}$ by Eq~(\ref{eq:grpo_loss}) using the frozen critic $Q_{\phi}$;
        \State Perform gradient descent to optimize $\phi$ and $\psi$;
        \State Update policy parameters $\theta_{\pi}$ (initialized from $\theta_{\pi}^{pre-trained}$);
        \EndFor
    \EndWhile
    
    \State Let $\theta_{\pi}^{\text{final}} \gets \theta_{\pi}$;
    \State \Return $\theta_{\pi}^{\text{final}}$.
\end{algorithmic}
\end{algorithm}

\begin{algorithm}[H]
\small
\caption{Bid Generation with SEGB}
\label{alg:online_compact_full}
\begin{algorithmic}[1]
\Require Trained planner $p_{\theta_{\text{LAD}}}$, evolved policy $\pi_\theta^{final}$, at each timestep $t$: state history $s_{\le t}$, action history $a_{<t}$, returns-to-go $R_{\le t}$.
\Ensure Bidding action $a_t$.

\State \textbf{Plan:} Predict next state: $\hat{s}_{t+1} \gets p_{\theta_{\text{LAD}}}(s_{\le t})$.
\State \textbf{Act:} Generate action: $a_t \sim \pi_\theta^{final}(a \mid s_{\le t}, a_{<t}, R_{\le t}, \hat{s}_{t+1})$.
\State \Return $a_t$.
\end{algorithmic}
\end{algorithm}
\vspace{-3mm}

\section{Experiments}
\vspace{-1mm}

In this section, we conduct extensive experiments to validate the effectiveness of our proposed SEGB framework. Our evaluation is designed to answer the following key research questions:
\begin{itemize}
    \item \textbf{RQ1: Overall Performance.} Does SEGB consistently outperform state-of-the-art auto-bidding baselines across different datasets and budget settings?
    \item \textbf{RQ2: Component-wise Contribution.} How do the core components of SEGB—specifically the LAD planner with foresight-awareness, and the offline policy evolution loop—each contribute to the final performance?
    \item \textbf{RQ3: Real-World Efficacy.} Can the performance gains observed in offline simulations translate into tangible business improvements in a large-scale, live production environment?
\end{itemize}

\subsection{Experimental Setup}
\subsubsection{Dataset} We adopt AuctionNet \cite{su2024auctionnet}, a large-scale simulated bidding benchmark by Alibaba. This dataset includes: (1) AuctionNet, with complete bidding trajectories, and (2) AuctionNet-Sparse, a sparser version with fewer conversions. Both contain ~500K trajectories from 10K episodes, each spanning 48 time steps. Detailed statistics are in Appendix Table~\ref{tab:statistics}.

\subsubsection{Evaluation Metrics}
Following AuctionNet, we use the score metric: $score = \sum_i (o_i v_i) \cdot \min \{(C/CPA)^\beta, 1\}, \beta=2$, balancing value maximization and KPI constraints. We employ rotation-based testing: each model replaces the 48 agents sequentially. For each rotation, we run 30 initializations and average the top-5 scores.
\subsubsection{Baselines}
We compare against representative baselines: \textbf{IQL}~\cite{Kostrikov2021iql} achieves policy iteration via expectile regression \cite{kostrikov2022offline} without evaluating out-of-distribution actions; \textbf{BCQ}~\cite{fujimoto2019off} constrains policies to behavior-close actions using VAE \cite{kingma2014auto}; \textbf{CQL}~\cite{kumar2020cql} learns conservative Q-functions by penalizing OOD actions to prevent overestimation; \textbf{DiffBid}~\cite{aigb} uses conditional diffusion models \cite{ho2020ddpm} for trajectory generation with inverse dynamics \cite{agrawal2016learning,pathak2018zero}; \textbf{DT}~\cite{chen2021dt} is a Transformer-based \cite{vaswani2017attention} sequence model; \textbf{GAS}~\cite{li2025gas} employs Monte Carlo Tree Search \cite{kocsis2006mcts} at inference.


\subsubsection{Implementation Details}
Experiments use PyTorch on NVIDIA A100 GPUs. Hyperparameters are in Appendix Table~\ref{tab:hyperparams}. Key configurations: LAD has 8 layers, 16 heads, 512-d embedding, $R=38$ diffusion steps, $\omega=0.2$ guidance, trained with AdamW (lr=$1 \times 10^{-5}$). Next-State-Aware DT has 6 layers, 8 heads, context=28, with LayerNorm and GELU. The IQL critic uses $\tau=0.8$ expectile, and GRPO uses $\beta=0.1$ KL penalty, both trained with lr=$3 \times 10^{-5}$.

\paragraph{Computational Efficiency.}
Offline training (LAD, NSA-DT, IQL, GRPO) takes ~4 hours on two A100 GPUs (one-time cost). Online inference achieves P99 latency <0.0375s, meeting the <100ms real-time constraint. GRPO adds zero online cost.

\subsection{Overall Performance Comparison (RQ1)}

Table~\ref{tab:overall} shows SEGB consistently outperforms all baselines across both datasets and all budgets. Key findings: 

(1) \textbf{SEGB achieves state-of-the-art performance} in all settings, with improvements ranging from 1.65\% to 12.25\% over the best baseline, demonstrating the overall superiority of our synergistic framework.

(2) \textbf{The performance gap widens on AuctionNet-Sparse}, validating that LAD's dense next-state prediction provides crucial guidance when long-term rewards are scarce. In sparse reward scenarios, the explicit short-term target from $\hat{s}_{t+1}$ becomes even more valuable.

(3) \textbf{DiffBid performs poorly}, generating entire trajectories globally. We attribute this to the difficulty of maintaining causal consistency (e.g., budget monotonicity) over long horizons. Its failure serves as strong evidence for our LAD design choice, which locally and autoregressively enforces these constraints.

\begin{table*}[t]
  \label{tab:overall}
  \centering
  \caption{\small Performance comparison on AuctionNet and AuctionNet-Sparse. SEGB consistently and significantly outperforms all baselines. Results are reported as mean score over 5 runs. The * indicates statistical significance (p < 0.05) over the best baseline.}
  \setlength\tabcolsep{9pt}
  \scalebox{0.8}{    \begin{tabular}{c|c|ccccccccc} 

    \toprule
    \toprule
    Dataset & Budget & BCQ & CQL & DiffBid & IQL  & DT  & GAS  & SEGB  & \textit{Improve} \\
    \midrule
    \multirow{5}[2]{*}{AuctionNet} & 50\%  & $100.22 \pm 2.96$    & $186.92 \pm 2.26$   & $33.25 \pm 1.52$   & $191.56 \pm 2.54$     & $191.36 \pm 3.73$   & \underline{$200.41 \pm 2.96$}   & \textbf{$203.71 \pm 1.35$*} & 1.65\% \\
          & 75\%  & $164.45 \pm 4.23$   & $256.35 \pm 5.10$   & $55.97 \pm 2.44$   & $260.68 \pm 4.67$     & $271.46 \pm 2.42$   & \underline{$279.31 \pm 5.32$}   & \textbf{$285.01 \pm 2.39$*} & 2.04\% \\
          & 100\% & $204.52 \pm 9.11$   & $318.11 \pm 2.70$   & $70.44 \pm 2.20$   & $325.89 \pm 4.16$     & $335.34 \pm 3.34$   & \underline{$347.07 \pm 3.34$}   & \textbf{$355.99 \pm 2.01$*} & 2.57\% \\
          & 125\% & $296.79 \pm 6.32$   & $374.10 \pm 3.00$   & $100.04 \pm 4.09$   & $377.37 \pm 3.92$     & $389.44 \pm 1.65$   & \underline{$398.12 \pm 5.88$}   & \textbf{$417.88 \pm 4.86$*} & 4.96\% \\
          & 150\% & $356.19 \pm 5.75$   & $420.47 \pm 5.40$   & $129.84 \pm 5.45$   & $421.46 \pm 4.63$     & $436.98 \pm 4.74$   & \underline{$449.78 \pm 5.50$}   & \textbf{$462.77 \pm 3.72$*} & 2.89\% \\
    \midrule
    \multirow{5}[2]{*}{AuctionNet-Sparse} & 50\%  & $15.50\pm0.88$   & $17.90\pm1.14$ & $5.67\pm0.39$  & $15.13\pm0.94$  & $18.94\pm0.99$   & \underline{$19.58\pm0.77$}  & \textbf{$20.82\pm0.79$*} & 6.33\% \\
          & 75\%  & $24.24\pm0.57$ & $25.36\pm0.77$ & $18.26\pm1.03$ & $7.15\pm0.64 $ & $25.89 \pm 0.93$  & \underline{$26.48 \pm 1.02$}  & \textbf{$28.58 \pm 1.00$*} & 7.93\% \\
          & 100\% & $30.88\pm0.69$ & $31.74\pm1.35$ & $9.34\pm0.57$  & $19.42\pm0.88$  & $32.57 \pm 1.05$  & \underline{$33.05 \pm 0.77$} & \textbf{$37.10 \pm 0.42$*} & 12.25\% \\
          & 125\% & $37.07\pm0.52$ & $37.69 \pm 0.83$  & $10.73\pm0.79$ & $20.68\pm1.18$   & $37.13 \pm 1.16$  & \underline{$37.87\pm0.82$}  & \textbf{$41.53 \pm 0.71$*} & 9.66\% \\
          & 150\% & $44.29\pm1.12$  & $43.02 \pm 0.58$  & $27.76\pm0.59$  & $21.62\pm1.85$    & $42.08 \pm 0.33$ & \underline{$44.42\pm1.91$}  & \textbf{$47.12 \pm 1.51$*} & 6.08\% \\
    \bottomrule
    \bottomrule
    \end{tabular}%
    }
  \label{tab:overall}%
\end{table*}%

\subsection{Ablation Study: Deconstructing SEGB's Success (RQ2)}
Table~\ref{tab:ablation} shows ablations on AuctionNet (100\% budget).

\begin{table}[t]
  \centering
  \caption{\small Ablation study on AuctionNet (100\% budget).}
  \begin{tabular}{llc}
    \toprule
    \textbf{Model Variant} & \textbf{Description} & \textbf{Score} \\
    \midrule
    \textbf{SEGB (Full)} & Our framework & \textbf{356.0} \\
    \quad w/o LAD & Diffusion + DT + GRPO & 349.3 \\
    \quad w/o $s'$ & DT + GRPO & 347.6 \\
    \quad w/o GRPO & LAD + DT & 346.4 \\
    \bottomrule
  \end{tabular}
  \label{tab:ablation}
\end{table}

Results reveal each component's critical contribution:

(1) \textbf{Removing GRPO} (-9.6 pts) confirms offline evolution's value. Without GRPO, the policy can only imitate the dataset; with it, the policy discovers superior strategies beyond what the offline data explicitly demonstrates.

(2) \textbf{Removing foresight} (-10.5 pts) shows explicit future state conditioning is critical. The predicted $\hat{s}_{t+1}$ provides a concrete, immediate target that RTG alone cannot offer.

(3) \textbf{Replacing LAD} (-14.5 pts) demonstrates LAD's causal planning is most crucial. Standard diffusion violates temporal constraints, while LAD's local autoregressive design ensures causally consistent trajectories.

All three components contribute synergistically to achieve state-of-the-art performance.

\subsection{Further Analysis: Efficacy of Offline Evolution}

\begin{figure}[t]
\centering
    \begin{subfigure}[b]{0.48\columnwidth}
        \includegraphics[width=\linewidth]{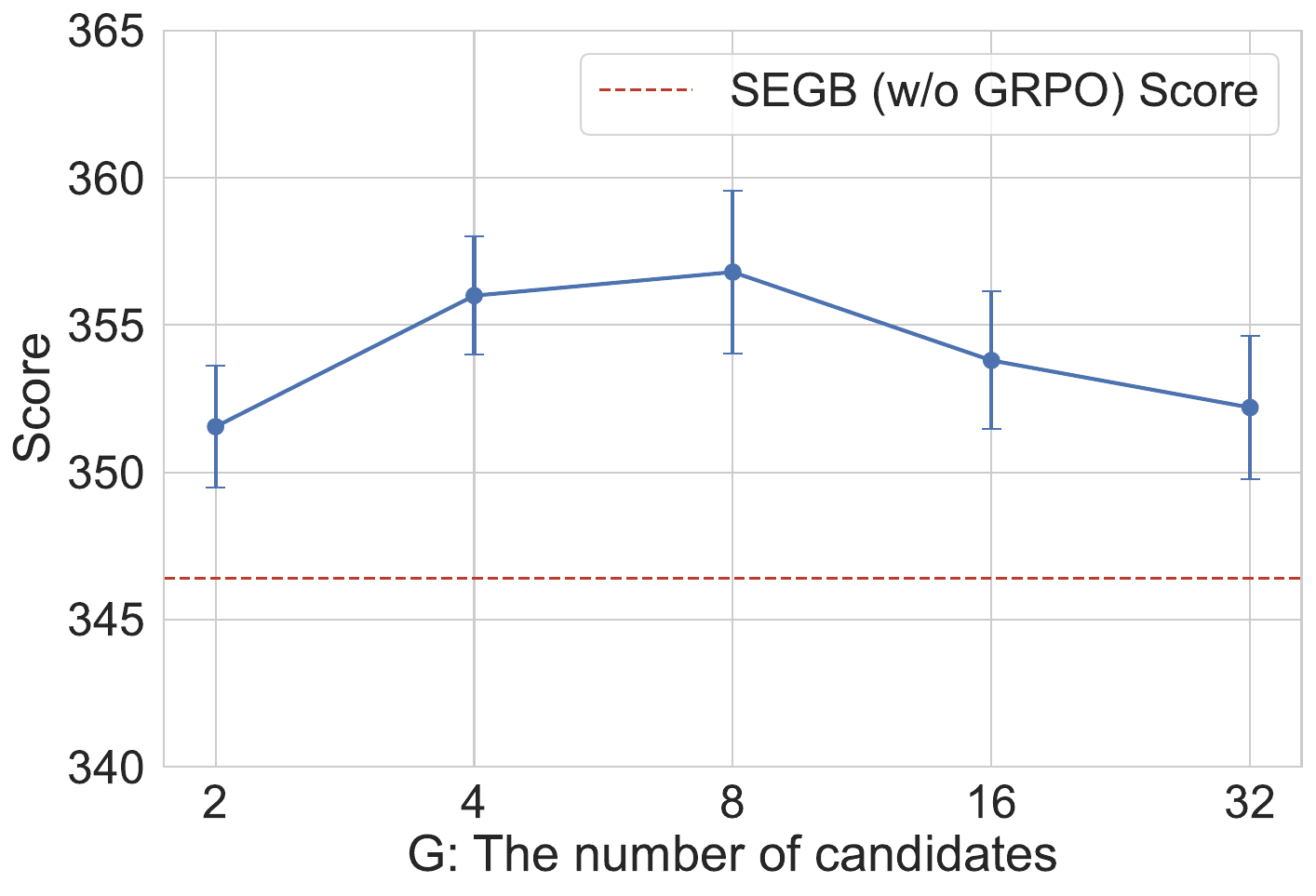}
        \caption{Impact of GRPO Group Size ($G$)} 
        \label{fig:grpo_g_size}
    \end{subfigure}
    \hfill
    \begin{subfigure}[b]{0.48\columnwidth}
        \includegraphics[width=\linewidth]{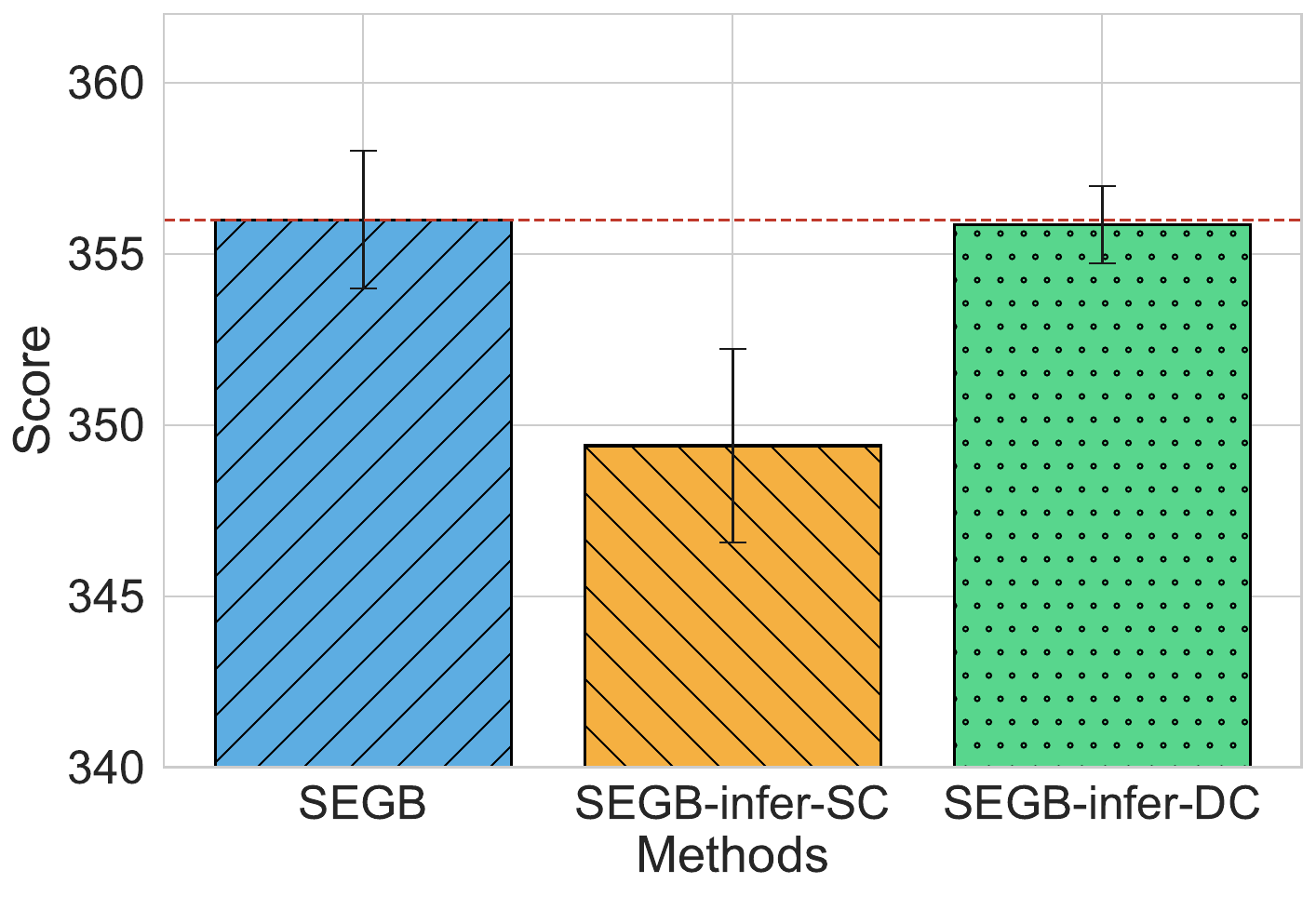}
        \caption{Impact of Online Voting ($K$)}
        \label{fig:voting_k}
    \end{subfigure}
    \caption{\small Further analysis on key hyperparameters.}
    \label{fig:further_analysis}
\end{figure}

To further understand SEGB's offline evolution, we analyze the impact of GRPO group size $G$ in Figure~\ref{fig:grpo_g_size}. The results show that $G=4$ achieves optimal performance. When $G$ is too small (e.g., 2), the policy suffers from insufficient exploration, limiting its ability to discover diverse strategies. When $G$ is too large (e.g., 8, 16), the advantage signal becomes diluted across too many samples, weakening the learning signal. $G=4$ strikes the best balance between exploration diversity and signal quality, enabling effective policy improvement.
\vspace{-2mm}

We further analyze the GRPO evolution stage by studying its sensitivity to the group size, $G$, and justifying our purely offline approach.

\paragraph{Impact of Group Size.}
The number of candidate actions, $G$, sampled in GRPO could influence the balance between performance and computational cost. To investigate this, we conduct a sensitivity analysis, with the results shown in Figure~\ref{fig:grpo_g_size}. We find that the performance is efficiently enhanced by increasing the number of candidates, with all settings significantly outperforming the baseline without GRPO. Specifically, the score rises sharply to 355.8 at $G=4$, capturing the vast majority of potential gains. While performance peaks at $G=8$, the minor improvement does not justify doubling the computational cost. Thus, we chose  $G=4$ as the optimal trade-off between performance and efficiency.

\paragraph{Sufficiency of Offline Evolution.}
Counter-intuitively, augmenting SEGB with a common online exploration technique—critic-based voting—proves detrimental. As visualized in Figure~\ref{fig:online_voting}, our final, purely offline model significantly outperforms both online voting variants. The performance collapses when using the same critic as GRPO ("Consistent Critic"), suggesting that online exploration merely disrupts an already-converged optimal policy. These findings provide strong evidence that SEGB's offline evolution is a sufficient and complete optimization process, validating our design of a fully self-contained framework.

\subsection{Online A/B Test (RQ3)}

To provide the ultimate validation and answer RQ3, we deployed SEGB in a large-scale online A/B test on the JD.com advertising platform.

\subsubsection{Experimental Setup}

\paragraph{Baseline and Deployment.}
Our baseline is the incumbent production model, a highly-optimized system rooted in Behavior Cloning (BC). This baseline was chosen for its exceptional stability and predictability—critical requirements for a platform handling hundreds of billions of daily requests. It represents a strong industrial benchmark that has been refined over years of production use. 

SEGB is deployed in a near-line serving system with GPU acceleration. The system consists of the LAD planner for future state prediction and the GRPO-refined policy for action generation (Algorithm~\ref{alg:online_compact_full}). The model is served via a distributed inference framework that ensures fault tolerance and load balancing across multiple GPU instances.

\paragraph{Multi-Stage Experiment Design.}
To ensure safe and robust validation, we conducted a multi-stage experiment:
\begin{itemize}
    \item \textbf{Phase 1 (Observation)}: May 28 - June 19, 2025. 
    SEGB was deployed on 20\% of budget and traffic to validate stability 
    and initial performance in a controlled setting.
    \item \textbf{Phase 2 (Scale-up)}: June 20 - June 30, 2025. 
    Following successful Phase 1 results, we scaled to 50\% of budget 
    and traffic for broader validation.
\end{itemize}
Consistent performance gains were observed across both phases, confirming 
SEGB's robustness to varying traffic conditions and budget allocations.

\paragraph{Latency and Performance.}
The deployed SEGB model achieves a P99 latency of under 0.0375s per 
request, compared to the baseline's ~0.015s. While SEGB introduces 
additional computational overhead due to the LAD planning step, this 
latency comfortably meets the platform's stringent <100ms constraint. 
Notably, GRPO incurs zero online computational cost, as it only refines 
the policy offline during training.

\subsubsection{Results and Business Impact}

As shown in Table~\ref{tab:online}, SEGB delivered substantial business 
impact, achieving a \textbf{+10.19\%} increase in target cost. This 
demonstrates that our synergistic, multi-stage architecture successfully 
bridges the critical offline-to-online gap. It proves that the intelligence 
learned and refined by SEGB's offline pipeline can generalize and excel 
in the dynamic, real-world online environment, delivering tangible value.

\begin{table}[t]
  \label{tab:online}
  \centering
  \caption{\small Online A/B test.}
  \scalebox{0.9}{
\begin{tabular}{cccc}
\hline
    Cost & Conversion & ROI & Target Cost \\ \hline
    +15.32\%       &  +8.13\%      &  +3.26\%      &  +10.19\%    \\ \hline
\end{tabular}
}
\label{tab:online}
\end{table}

\subsubsection{Robustness to Distribution Shift}

A critical concern in offline RL is the ability to generalize beyond the training distribution. To validate SEGB's out-of-distribution (OOD) robustness, we analyze performance across natural distribution shifts encountered in production.

\paragraph{Traffic Distribution Shifts.}
The consistent performance gains observed across our multi-stage experiment (from 20\% to 50\% of traffic) provide evidence of robustness to traffic distribution changes. The +10.19\% target cost improvement was stable across both phases, despite significant differences in traffic volume, user composition, and competitive dynamics.

\paragraph{Cold-Start Generalization.}
More critically, we evaluate SEGB on cold-start advertising campaigns—a true OOD test where the system has no historical data for the specific campaign. This scenario is particularly challenging as it requires the policy to generalize based solely on learned bidding principles, without campaign-specific fine-tuning. On a held-out set of 500+ cold-start campaigns, SEGB achieved a +18.03\% increase in target cost compared to the baseline, significantly exceeding the +10.19\% average improvement. This remarkable performance confirms SEGB's strong generalization capability and validates the effectiveness of our offline evolution approach for handling distribution shift.

\section{Related Work}

Our work is situated at the intersection of auto-bidding, offline reinforcement learning, and generative modeling. In this section, we review the most relevant literature in these areas and contextualize the contributions of SEGB.

\noindent\textbf{Auto-Bidding as a Sequential Decision Problem} Auto-bidding aims to optimize advertisers' strategies by algorithmically balancing budget constraints, return on investment (ROI), and campaign objectives \cite{zhang2014optimal, zhang2016feedback}. Early approaches often relied on control theory, such as PID controllers \cite{chen2011pid}, to manage budget pacing. While simple and effective, these methods lack the ability to adapt to highly dynamic traffic environments. Subsequent methods based on online linear programming \cite{agarwal2014budget} showed promise but were sensitive to the accuracy of traffic predictions.

Recognizing the need for adaptive strategies, the community increasingly formulated auto-bidding as a sequential decision-making problem \cite{jin2018mulagent}. This framing paved the way for data-driven solutions, particularly those based on reinforcement learning, which we discuss next.

\noindent\textbf{Reinforcement Learning for Auto-Bidding} Reinforcement Learning (RL) formulates bidding as a process where an agent learns to make optimal decisions to maximize long-term rewards. Early notable works applied DRL to the single-agent bidding problem \cite{zhao2018deep} or explored multi-agent competition dynamics \cite{jin2018mulagent}.

Given the high cost and risk of online exploration in advertising, most modern systems operate in the \textbf{offline RL} setting \cite{levine2020offline}, learning policies from large, static datasets. A central challenge in offline RL is the \textbf{distributional shift} problem. To mitigate this, numerous algorithms have emerged. One category focuses on policy constraints, ensuring the learned policy stays close to the data-collection policy, as exemplified by BCQ \cite{fujimoto2019off}. Another category, known as policy regularization, penalizes the value estimates of out-of-distribution (OOD) actions. This includes Conservative Q-Learning (CQL) \cite{kumar2020cql} and methods that explicitly quantify uncertainty, such as EDAC, which uses ensembles to create a pessimistic value function \cite{an2021uncertainty}. A third approach, which we adopt for our critic training, is to implicitly avoid OOD evaluation altogether. Implicit Q-Learning (IQL) \cite{Kostrikov2021iql} achieves this through expectile regression \cite{koenker2001quantile}, offering a robust method for value estimation from offline data.

Limitations and Our Approach: While powerful, these value-based offline RL methods often rely on the Markov assumption. However, studies like AIGB \cite{aigb} have shown that bidding environments are often non-Markovian. This limitation motivates a shift towards sequence modeling, where decisions are based on extended histories rather than just the current state.

\noindent\textbf{Generative Approaches for Offline RL and Bidding} Generative models have achieved remarkable breakthroughs in modeling complex data distributions. Early successes were established by Variational Autoencoders (VAEs)~\cite{vae} and Generative Adversarial Networks (GANs)~\cite{gan}. More recently, Denoising Diffusion Probabilistic Models (DDPMs)~\cite{ho2020ddpm} have emerged as the state-of-the-art for high-fidelity synthesis, demonstrating unprecedented quality in domains like image and audio generation~\cite{kong2020diffwave}. This wave of innovation has naturally inspired new ways of tackling sequential decision-making problems.
In offline RL, this led to a paradigm shift treating the problem as sequence modeling, powered by the Transformer architecture~\cite{vaswani2017attention}. This was pioneered by works like Decision Transformer (DT)~\cite{chen2021dt}, which directly models trajectories to generate actions conditioned on a desired return. This paradigm was quickly adapted to auto-bidding to handle its non-Markovian nature, leading to several powerful but ultimately incomplete approaches.

These pioneering generative methods, however, have their own limitations which SEGB is designed to systematically address:
\begin{itemize}
    \item \textbf{Lack of Causal Adherence:} Inspired by the success of Denoising Diffusion Probabilistic Models (DDPMs)~\cite{ho2020ddpm}, DiffBid~\cite{aigb} proposed using a conditional diffusion model to generate trajectories. However, generating the entire sequence in a "global" fashion struggles to enforce real-world causal constraints (e.g., budget monotonicity). \textbf{SEGB addresses this with our Local Autoregressive Diffusion (LAD) model}, which enforces causality by generating states sequentially, a philosophy aligned with recent advances in continuous-space autoregressive generation~\cite{localdiff}.

    \item \textbf{Reactive vs. Proactive Decision-Making:} Standard DT and its variants, such as GAS~\cite{li2025gas}, are purely \textbf{reactive}, conditioning only on past events and a final goal. They lack a mechanism for short-term, tactical planning. \textbf{SEGB introduces the Next-State-Aware DT}, which incorporates explicit foresight from our LAD planner, enabling proactive decision-making.

    \item \textbf{Static Policy vs. Evolving Policy:} While methods like GAS introduce an online search component, their underlying policy model remains \textbf{static} after offline training, unable to discover strategies superior to those in the dataset. \textbf{SEGB proposes a complete offline policy evolution loop}, where the policy is improved offline with GRPO~\cite{grpo}—an offline policy gradient method chosen for its stability—enabling the discovery of superior strategies entirely from static data.
\end{itemize}

In summary, SEGB advances the state of the art by creating a synergistic framework that combines the strengths of generative planning and offline policy evolution to create a more robust, proactive, and adaptive bidding agent.

Recently, VACO \cite{vaco} introduced bi-level optimization for value-aligned imitation, and A2PO \cite{a2po} proposed generative action modeling conditioned on advantages. In contrast, SEGB's Local Autoregressive Diffusion (LAD) focuses on high-fidelity state planning to provide explicit foresight, which we argue is more robust for non-Markovian bidding dynamics.

In summary, SEGB advances the state of the art by creating a synergistic framework that combines the strengths of generative planning and offline policy evolution to create a more robust, proactive, and adaptive bidding agent.

\section{Conclusion}

In this paper, we addressed the critical challenge of bridging the offline-to-online gap in automated bidding by proposing the Self-Evolved Generative Bidding (SEGB) framework. Our synergistic architecture systematically tackles the planning deficits and exploration dilemmas of offline learning. It uniquely integrates a Local Autoregressive Diffusion (LAD) model for causally-consistent state planning, a Next-State-Aware Decision Transformer for foresight-driven action generation, and an offline policy evolution loop featuring GRPO fine-tuning.

Experimental results demonstrated SEGB's state-of-the-art performance on public benchmarks. Crucially, our approach was validated in a large-scale online A/B test, where it delivered a significant +10.19\% increase in target cost, proving its ability to translate offline-learned policies into real-world business value.

While SEGB is effective, future work could focus on enhancing its robustness to non-stationary market dynamics and exploring more computationally efficient planning models. In conclusion, SEGB not only presents a powerful new solution for auto-bidding but also offers a promising blueprint—integrating planning, foresight, and evolution—for other complex sequential decision-making problems.


\appendix

\section{Dataset Statistics}

Table~\ref{tab:statistics} summarizes the key statistics of the AuctionNet and AuctionNet-Sparse datasets used in our experiments.

\begin{table}[h]
  \centering
  \caption{Data statistics for AuctionNet and AuctionNet-Sparse.}
  \begin{tabular}{lcc}
    \toprule
    \textbf{Params} & \textbf{AuctionNet} & \textbf{AuctionNet-Sparse} \\
    \midrule
    Trajectories & 479,376 & 479,376 \\
    Delivery Periods & 9,987 & 9,987 \\
    Time steps per trajectory & 48 & 48 \\
    State dimension & 16 & 16 \\
    Action dimension & 1 & 1 \\
    Return-To-Go dimension & 1 & 1 \\
    Action range & [0, 493] & [0, 589] \\
    Impression value range & [0, 1] & [0, 1] \\
    CPA range & [6, 12] & [60, 130] \\
    Total conversion range & [0, 1512] & [0, 57] \\
    \bottomrule
  \end{tabular}
  \label{tab:statistics}
\end{table}

\section{Hyperparameter Details}

Table~\ref{tab:hyperparams} provides the complete hyperparameter configuration for all SEGB components.

\begin{table}[h]
  \centering
  \caption{Detailed hyperparameters for SEGB components.}
  \label{tab:hyperparams}
  \footnotesize
  \begin{tabular}{llc}
    \toprule
    \textbf{Component} & \textbf{Hyperparameter} & \textbf{Value} \\
    \midrule
    \multirow{8}{*}{\textbf{LAD Planner}} & Layers & 8 \\
    & Attention Heads & 16 \\
    & Embedding Dimension & 512 \\
    & Context Length & 48 \\
    & Diffusion Steps ($R$) & 38 \\
    & Guidance Strength ($\omega$) & 0.2 \\
    & Conditional Dropout Rate & 0.2 \\
    & Noise Schedule ($\beta_k$) & Linear, $10^{-4}$ to $0.02$ \\
    \midrule
    \multirow{7}{*}{\textbf{Next-State-Aware DT}} & Layers & 6 \\
    & Attention Heads & 8 \\
    & Embedding Dimension & 512 \\
    & Context Length & 28 \\
    & Dropout (Attn, Embed) & 0.1 \\
    & Activation Function & GELU \\
    & Pre-training Batch Size & 256 \\
    \midrule
    \multirow{6}{*}{\parbox{3.5cm}{\textbf{IQL Critic \& GRPO Post-training}}} & Critic Architecture & Same as DT \\
    & IQL Expectile ($\tau$) & 0.8 \\
    & Discount Factor ($\gamma$) & 0.99 \\
    & GRPO KL Penalty ($\beta$) & 0.1 \\
    & GRPO Clipping ($\epsilon$) & 0.1 \\
    & GRPO Group Size ($G$) & 4 \\
    \midrule
    \multirow{5}{*}{\textbf{General Optimization}} & Optimizer & AdamW \\
    & Learning Rate (LAD) & $1 \times 10^{-5}$ \\
    & Learning Rate (DT, IQL) & $3 \times 10^{-5}$ \\
    & Weight Decay & 0.01 \\
    & Gradient Clipping Norm & 1.0 \\
    \bottomrule
  \end{tabular}
\end{table}

\end{document}